\documentclass[journal]{IEEEtran}

\usepackage{amssymb}
\usepackage{amsmath}
\usepackage{amsthm}
\usepackage{amsfonts}
\usepackage{slashed}

\usepackage[table,xcdraw]{xcolor}
\usepackage{hyperref}
\usepackage{listings}
\usepackage{tabularray}
\usepackage{siunitx}
\usepackage{tablefootnote}
\usepackage[export]{adjustbox}

\newcounter{bla}

\newtheorem{proposition}{Proposition}
\newtheorem{exmp}{Example}[section]

\newcommand\restr[2]{{
  \left.\kern-\nulldelimiterspace 
  #1 
  \littletaller 
  \right|_{#2} 
  }}

\newcommand{\littletaller}{\mathchoice{\vphantom{\big|}}{}{}{}}

\lstdefinelanguage{Julia}%
  {morekeywords={abstract,begin,break,case,catch,const,continue,do,else,elseif,%
      end,export,false,for,function,immutable,import,importall,if,in,%
      macro,module,otherwise,quote,return,switch,true,try,type,typealias,%
      using,while},%
   sensitive=true,%
   alsoother={\$},%
   morecomment=[l]\#,%
   morecomment=[n]{\#=}{=\#},%
   morestring=[s]{"}{"},%
   morestring=[m]{'}{'},%
   basicstyle=\ttfamily\small,%
   breaklines=true,%
}[keywords,comments,strings]%

\begin{document}

\title{Optimizations on Graph-Level for Domain Specific Computations in Julia and Application to QED}

\author{\IEEEauthorblockN{
A.~Reinhard\IEEEauthorrefmark{1}\IEEEauthorrefmark{2}\IEEEauthorrefmark{3},
S.~Ehrig\IEEEauthorrefmark{1}\IEEEauthorrefmark{2},
R.~Widera\IEEEauthorrefmark{1},
M.~Bussmann\IEEEauthorrefmark{2},
U.~Hernandez~Acosta\IEEEauthorrefmark{1}\IEEEauthorrefmark{2}}\\
\IEEEauthorblockA{
\IEEEauthorrefmark{1} Helmholtz-Zentrum Dresden-Rossendorf, Bautzner Landstraße 400, 01328 Dresden\\
\IEEEauthorrefmark{2} Center for Advanced Systems Understanding, Untermarkt 20, 02826 Görlitz\\
Email: \IEEEauthorrefmark{3} a.reinhard-science@proton.me}}


\maketitle

\begin{abstract}
Complex computational problems in science often consist of smaller parts that can have largely distinct compute requirements from one another. For optimal efficiency, analyzing each subtask and scheduling it on the best-suited hardware would be necessary. Other considerations must be taken into account, too, such as parallelism, dependencies between different subtasks, and data transfer speeds between devices. To achieve this, directed acyclic graphs are often employed to represent these problems and enable utilizing as much hardware as possible on a given machine. In this paper, we present a software framework written in Julia capable of automatically and dynamically producing statically scheduled and compiled code. We lay theoretical foundations and add domain-specific information about the computation to the existing concepts of DAG scheduling, enabling optimizations that would otherwise be impossible. To illustrate the theory we implement an example application: the computation of matrix elements for scattering processes with many external particles in quantum electrodynamics.\\


\end{abstract}

\section{Introduction}
\IEEEPARstart{O}{ptimizing} the performance of high-performance code is challenging because many factors influence it. On the hardware side, these include caching, thread counts, and hyper-threading abilities, the available microarchitecture, hardware accelerators, and the data interconnection speed between CPUs and the accelerators \cite{hardware_review}. However, on the tooling side, many performance-relevant choices also exist, such as compiler options or libraries for low-level operations and GPU usage \cite{kokkos, alpaka}. Lastly, software choices such as memory layouts \cite{llama}, design, and language influence the performance. The combination of the already large optimization space on the software side and the wide variety of hardware makes it very difficult to solve problems efficiently while maintaining portability (especially performance-portability) at the same time.\\
Compilers are very good at optimizing small pieces of code to near-optimal levels, but are constrained by guarantees and a lack of domain knowledge about what is being computed. Therefore, higher-level optimizations are often out of scope for compilers because they rely on properties that can not generally be assumed, or because they involve too large parts of the code base. Some domain-specific languages (DSL) get around this by essentially hard-coding paths of optimization into their design. However, by their nature, they are not portable to other domains and require large amounts of implementation work on their own.\\
In HPC systems and software, multiple layers of optimization are commonly available. The two polar ends of the spectrum are compilers, which operate at the lowest level of abstraction, i.e., closest to the physical hardware, and frameworks like MPI \cite{openmpi,mpi_reference}, which operate at a very high level of abstraction. Engines like this can target large clusters and parallelize effectively, but are unable to optimize the program flow itself because they lack meta-information about what is being executed. Furthermore, for workflow engines, the relevant task kernels are often complex and comparatively long-running themselves, making the overhead of dynamic scheduling negligible in most cases.\\
To reconcile these two extremes, we propose an approach representing computational problems as DAGs, made up of nodes carrying compute kernels to allow the targeting of large compute systems. On top of this well-known design, we propose an optimization scheme on the DAG level by analyzing and muting the DAG structure. The required domain knowledge about the problem is implicitly given by the DAG itself, and can be further fine-tuned by assigning attributes to nodes to allow term rewriting. Importantly, this approach does not aim to replace the optimizations of compilers, but work with and benefit them.\\
Frameworks and libraries targeting similar goals to ours have been proposed before. StarPU is a runtime system to dynamically schedule tasks on heterogeneous systems, aiming to provide a unified programming interface for multi-core and GPU programming \cite{starpu_1}. However, it still requires separate implementations of the tasks themselves to be provided for all target architectures. It also does not optimize the graph on its own. Another is SuperGlue, which uses a stronger but similar version of a DAG representation to encode both data dependencies and concurrency exclusions \cite{superglue}. It uses a fully dynamic scheduling approach to scale problems efficiently to many CPU cores.\\
Much previous work has also been done on scheduling DAG representations of large computing jobs. We currently target heterogeneous single-node devices, such as a typical HPC cluster node with \num{2} CPU cores and one or several GPUs. However, most research in this domain targets either low-level scheduling in the backends of compilers \cite{lee_compiler_2000, kessler_scheduling_1998, wang_compiler-assisted_2010, rotem_glow_2019}, or scheduling for large multi-node programs \cite{braun_comparison_1999}. In contrast, the tasks we schedule are large compared to the ones a typical compiler schedules, but small for a typical high-performance cluster. For example, a task in our framework might be a small matrix multiplication, while the compiler would see the underlying floating-point additions and multiplications. On the other hand, a framework at a higher level of abstraction might see an entire timestep of a simulation as one task.\\
Furthermore, an abundance of scheduling algorithms exist to solve or approximate solutions to scheduling problems. Some target single-threaded execution \cite{kessler_scheduling_1998}, some target multi-threaded but homogeneous execution \cite{liou_comparison_1997, el-rewini_task_1995, kwok_benchmarking_1998, chen_case_2022}, and some large-scale heterogeneous systems \cite{topcuoglu_performance-effective_2002, braun_comparison_1999, hagras_near_2004, wang_hsip_2016, khogali_cost_2013, sohani_predictive_2021, khan_task_2017}.\\
Categorization also exists between dynamic and static scheduling, where dynamic schedulers handle graphs whose topology or weights may change during execution \cite{wu_partitioning_2018, sohani_predictive_2021, qasim_dynamic_2017}, in contrast to static schedulers, which can rely on an unchanging graph \cite{kwok_static_1999, braun_comparison_1999}.\\
To solve these many constraints for problems with often intractable optimal solutions \cite{el-rewini_task_1995}, neural-net-based algorithms have been proposed in recent years, both for scheduling and also optimization \cite{geng_profile-based_2022}.\\
Since our proposed structure is based on taking advantage of domain-specific knowledge, it is also closely related to domain-specific languages.
\subsection{Choice of Language}
To implement our approach, some requirements must be met by the underlying infrastructure, first and foremost the language.\\
For easy adoption and quick implementations and subsequent extensions, the language of choice should provide a high degree of expressiveness. The barrier-of-entry and time-to-first-simulation should be as low as possible. Next, to implement the translation of a graph-level representation to executable machine code, the language must provide extensive capabilities for meta-programming and code generation. For ease-of-use it is also beneficial if no switch of tools or language is necessary to run the generated code. Furthermore, the language should be able to target various architectures, most importantly GPUs of the main vendors. In the best case, the user can write completely hardware agnostic code that runs seamlessly on any given hardware. Since we want to target scientific computing and high performance codes, the performance of the language must also be high and portable to the target architectures.\\
Choosing Julia as the language for the implementation of this project fulfills most of these requirements. Julia's powerful methods of meta-programming and its ability to understand and represent its own code are very well-suited for the code generation steps \cite{julia_metaprogramming}. Julia also allows us to generate and call a function without leaving the language or the running session. Enabling behavior like this, while possible, would be cumbersome in compiled languages such as C/C++ and slow in fully interpreted languages like Python. Importantly, this allows us to fully compile and execute a statically created schedule, benchmark it, without being locked in to this schedule. This allows us to use complex and compute intensive scheduling algorithms, because their overhead is not included in the runtime, and optimize with near perfect knowledge of the effect on the final runtime.\\
Furthermore, we can use packages like \texttt{KernelAbstractions.jl} \cite{KernelAbstractions} to generate code for many architectures seamlessly without much user effort, i.e., without requiring the user to provide extra implementations specific accelerators. This enables us to generate code for single- or multi-threaded CPUs, GPUs of multiple vendors (e.g., AMD, NVIDIA, Intel, Apple) \cite{julia_bridging_hpc}, and, in the future, even FPGAs \cite{julia_fpga}.\\
Finally, Julia is a beginner-friendly language \cite{why_julia} while still enabling highly-optimized and scalable code to be written by experts \cite{julia_bridging_hpc, julia_on_gpu}. This keeps the barrier of entry low for new developers or scientists from various non-computer-science domains while allowing low-level optimizations where necessary.\\
\subsection{CDAG Representation Requirements}
Representing a function as a DAG imposes some restrictions, so it is not entirely universally applicable. The first and most rigid requirement is that we assume that the DAG is static, i.e., it can not dynamically change its structure during execution. This is necessary to allow static analysis, optimization, and compilation of the DAG. In practice this means that a generator must exist that produces the complete CDAG with set nodes, their functions, defined inputs and outputs, and edges between the nodes. This generator is user provided and specific to the application, but the subsequent analysis, optimization and compilation is domain-agnostic.\\
Some soft requirements apply to the computations of individual nodes of the computable DAG. These computations are functions mapping from any number and type of arguments to a result. This result, too, can have any size and type, so it could, for example, be an array or tuple. However, any subsequent computation will receive the whole result as is.\\
To reduce ambiguity, we call the computations in this context \textit{kernels}:
\begin{enumerate}
	\item The same kernels should occur in multiple nodes in a given DAG. If this is not the case, the DAG can still be computed but not optimized effectively since the optimization space is small.
	\item Kernels should be pure \footnote{A pure function is one that does not change or rely on the global state of the program and returns the same result for identical arguments every time.}, consistent \footnote{A consistent function's manner of termination and result are identical for identical parameters.}, and globally terminating functions \footnote{A globally terminating function is one that can be proven to always terminate eventually.}. If the language can infer these function attributes, it allows more optimizations and produces more efficient code.
	\item Kernels should have a mostly parameter-independent runtime. This allows the scheduler to make better decisions about where to schedule certain parts of the CDAG for an even load on the available hardware. Note that this still allows kernels to scale as long as the information affecting the runtime is known at compile time, for example, through type information of the arguments.
\end{enumerate}
\subsection{Domain-Specific Applications}\label{sec:domain_specific}
To exhibit the capabilities of \texttt{ComputableDAGs.jl}, we implemented several domain-specific applications solving problems of high complexity from various scientific fields. The first application is an implementation of perturbative quantum electrodynamics (QED), which is a quantum field theory (QFT) and known to be able to produce the most accurate predictions in all of science \cite{Gabrielse:2006gg}. In the context of \texttt{ComputableDAGs.jl}, computations of scattering processes in QED are prime examples for problem statements with rapidly growing complexity; often stronger than factorial. These computations already have an inherent graph structure (the Feynman diagrams, see below), which can grow to arbitrary size and complexity, depending on the particles scattered. However, despite the complicated structure, the fundamental building blocks can be reduced to simple compute tasks like very small matrix multiplications of just $4\times 4$ matrices.  

One of the central quantities to be calculated in any quantum field theory is the scattering matrix element (S-matrix for short), which is a probabilistic measure of particles being scattered according to a given scattering process. The calculation of these S-matrices is the most compute-intensive part of Monte-Carlo event generation, and, therefore, a target where performance improvements could be highly beneficial.
We also briefly investigated a non-physical QFT named the "ABC model" \cite{introparticlephysics}, which is primarily used for educational purposes. It retains the structural complexity of other QFTs, but contains much less computationally intensive kernels. This makes it an interesting comparison to our implementation of QED, where the atomic building blocks we use are much more compute-intensive. We use this to investigate the impact of the amount of computation within the fundamental compute tasks on the overall performance and optimization capabilities of our domain-specific DAG-representation. 

As a second showcase for the applicability of \texttt{ComputableDAGs.jl}, we have also provided an implementation of the Strassen matrix multiplication algorithm \cite{strassen} using the package, which can be found in Appendix \ref{ch:strassen}.
\subsection{Quantum Electrodynamics}\label{sec:qed}
\begin{figure}[t]
    \centering
    \begin{align*}
    \includegraphics[width=0.45\linewidth,valign=c]{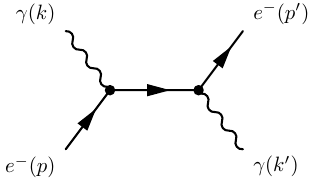} +
    \includegraphics[width=0.45\linewidth,valign=c]{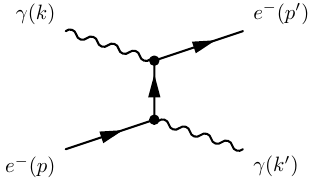} 
    \end{align*}
    
    \caption{The two Feynman diagrams contributing to the Compton scattering process $e^- \gamma \to e^- \gamma$ at tree-level. In the left diagram, the incoming electron first interacts with the incoming photon. The resulting inner line (also called a virtual particle) then interacts with the outgoing photon and outgoing electron. In the right diagram, the order of the photon interactions is reversed.}
    \label{fig:ex_compton}
\end{figure}

In the following section, we give a very brief overview to the computational method we implemented to calculate S-matrices in QED. For a more detailed introduction, we refer the the reader to the extensive standard literature (see, e.g., \cite{introparticlephysics,Peskin:1995ev,srednicki2007quantum}).
Quantum electrodynamics describes the interaction of fermions (electrons, muons, tauons, and their anti-particles) and photons including all relativistic and quantum mechanical phenomena.\\
In any QFT, such as QED, the scattering matrix element (S-matrix) is a probabilistic measure for a given group of particles to undergo a given scattering process. For small coupling constant, as in QED, where \(\alpha\approx 1/137\), the S-matrix can expand perturbatively in powers of the coupling. One well-established method to represent the contributions at each order are the Feynman diagrams, which are translated into mathematical expressions using Feynman rules. Figure \ref{fig:ex_compton} shows the tree-level Feynman diagrams for Compton scattering, \(e^-\gamma \to e^-\gamma\). The external solid lines denote the electron bispinors \(u, \bar u\), the internal line the electron propagator:  \(S(Q) = \frac{i(\slashed Q + m)}{Q^2 - m^2 + i\epsilon}\), where Q is the respective four-momentum transfer, we used Feynman’s slash notation  \(\slashed a = \gamma^\mu a_\mu\) with Dirac’s gamma matrices \(\gamma^\mu\), and we exploit Einstein’s sum convention. The wavy lines denote the photon polarization vectors \(\varepsilon, \varepsilon^{\prime}\), the dot connecting internal and external lines is the QED vertex, which contributes the factor \(-e\gamma^\mu\), where \(e\) is the elementary electric charge. A complete set for all Feynman rules, not only for QED, can be found in any standard introduction on quantum field theory (see, e.g.,  \cite{Peskin:1995ev}). Applying the Feynman rules and following the arrows on the solid lines in reverse order, yields the well-known tree-level matrix element of Compton scattering:
\begin{align}
\mathcal M = (-ie)\left[
	\bar u(p^{\prime}) \slashed\varepsilon^{\prime *} S(Q)\slashed\varepsilon u(p) + \bar u(p^{\prime}) \slashed\varepsilon S(Q’)\slashed\varepsilon^{\prime *} u(p)
\right], 
\end{align}
where \(Q = p+k\) and \(Q^{\prime} = p-k^{\prime}\) are the respective momentum transfers.
The tree-like topology of the considered Feynman diagrams makes their computation a natural application for CDAG-based implementations. While the number of Feynman diagrams for general scattering processes grows superfactorially with the number of scattered particles, the number of possible momentum combinations just grows exponentially. This opens up significant opportunities for optimizations of CDAG-based implementations.\\
Considerably, reusing intermediate results can significantly reduce the compute time, especially if one considers the sum over different spins and polarizations, where the same Feynman diagram must be evaluated repeatedly for many combinations of inputs.\\ 
Moreover, the entire topology of the CDAG representing the evaluation of all Feynman diagrams for a given scattering process is statically known, satisfying the hard technical requirement for the CDAG representation. Furthermore, the same compute kernels appear many times across the CDAG, all the kernels are pure, i.e., do not change any external states of the program, and all except \texttt{sum} run in constant time. Therefore, all soft requirements for a problem to be efficiently usable as a CDAG are met by the Feynman diagram technique.

\section{Computable DAGs}\label{ch:computable_dags}
\begin{figure}[t]
    \centering
    \includegraphics[width=0.55\linewidth]{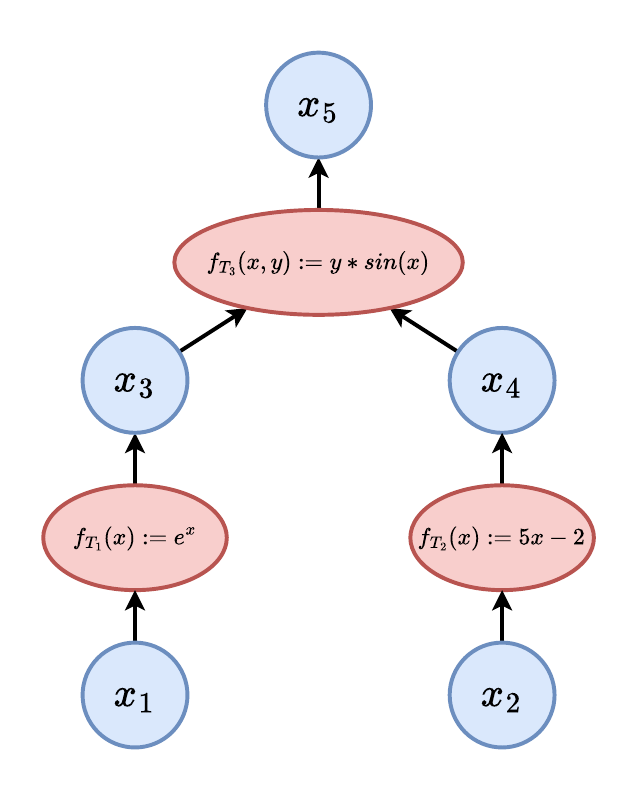}
    \caption{Example of a Computable DAG with functions assigned to every compute node. Data nodes are shown in blue,  compute nodes in red. The CDAG has two real numbers, $x_1$ and $x_2$, as inputs and another real number, $x_5$, as output. By following along the edges, we can find that the intermediate value $x_3 := e^{x_1}$ and $x_4 := 5x_2 - 2$. Putting these results into the third data node, we find that the result $x_5 := x_4 * \sin(x_3) = (5x_2 - 2) * \sin(e^{x_1})$. This is the function the CDAG computes.}
    \label{fig:ex_cdag}
\end{figure}

A directed graph is a tuple of a set of vertices and directed edges where each directed edge is a tuple of the two vertices that it points from and to. A path of non-zero length along directed edges in a graph such that the first and last vertex of the path are the same is called a cycle. If no such cycle exists within a given graph, the graph is called acyclic, and thus a directed acyclic graph, or DAG.\\
We can use a DAG to represent a computation by associating each graph vertex with a function, which we call \textit{task}. The DAG's edges represent dependencies between the functions associated with different vertices and do not carry any functionality by themselves.\\
To allow for a more accurate representation that helps with scheduling and execution, we introduce a distinction between \textit{compute tasks} and \textit{data tasks}. Compute tasks are functions that take any non-zero number and any type of parameters and produce a single result. Therefore, the associated node is a \textit{compute node} with any number of unique incoming edges but only one unique outgoing edge \footnote{It is allowed that a compute node has multiple outgoing edges, however, they will all represent the same single result. This is useful as an intermediate step in graph-muting optimizations.}. Data tasks are identity functions but can duplicate their outputs. Therefore, the associated \textit{data node} has exactly one incoming edge but any number of outgoing edges. Special types of data nodes can have only an outgoing or only an incoming edge, making them an input or output of the entire DAG. These nodes represent the cost of copy, load or store operations on the underlying hardware.\\
The duality of compute and data nodes restricts the allowed edges to those that connect a data node and a compute node in either direction. Edges between two data or two compute nodes are disallowed.\\
We call a DAG following these definitions a \textit{Computable DAG}, or \textit{CDAG}. A CDAG represents a computation, for example, as shown in Figure \ref{fig:ex_cdag}. In this figure, there are three individual compute nodes $T_1$, $T_2$, and $T_3$ with distinct kernels. $x_1$ is transformed by $T_1$ into $x_3$, and likewise $x_2$ is transformed by $T_2$ into $x_4$. $T_3$ takes two arguments and calculates the example's final result $x_5$ from $x_3$ and $x_4$.\\
To implement the theory we described, we have developed the Julia package \href{https://github.com/ComputableDAGs/ComputableDAGs.jl}{\texttt{ComputableDAGs.jl}} \cite{computabledagsjl}.\\
\begin{figure}[t]
    \centering
    \includegraphics[width=\linewidth]{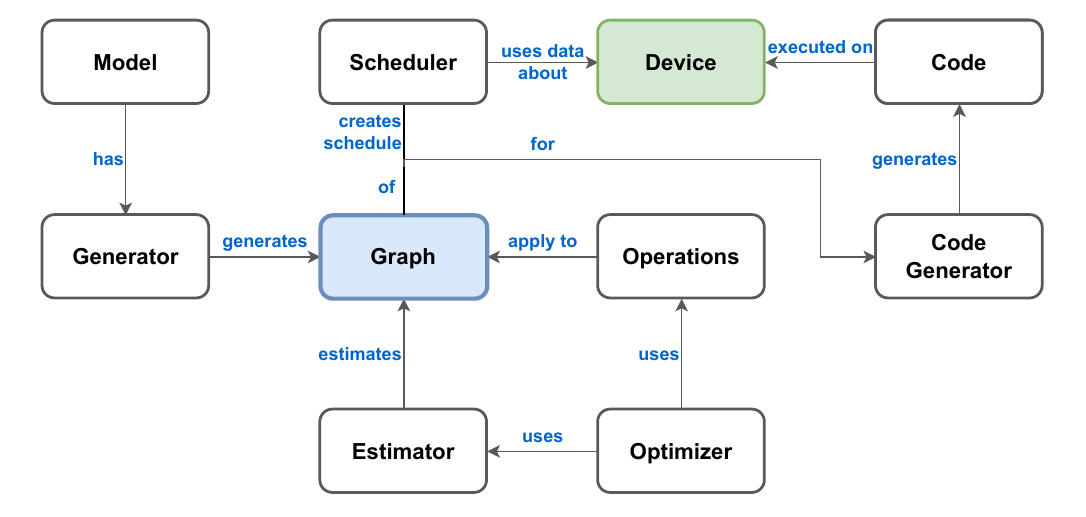}
    \caption{Modules of the software and their interactions}
    \label{fig:overview}
\end{figure}
\subsection{Design}
\texttt{ComputableDAGs.jl} consists of many interacting parts consisting of mostly orthogonal interfaces, simplifying the extension of one part with new functionality to an implementation of the respective interface.\\
The overall structure is depicted in \ref{fig:overview}. The entry point to the functionality consists of a \textit{model} with its respective \textit{generator}. The model represents a domain-specific problem space. For a specific instance in this problem space, a user-provided generator generates a CDAG, the central object of the package. For example, in our own project application, as discussed in more detail in Section \ref{sec:domain_specific} above, the model is perturbative QED, and a problem instance would be one specified particle scattering process. From the scattering process definition, a generator creates a CDAG representing the desired computation for this problem instance. The model and the generator are the only two parts necessary to implement a domain-specific application. Note that the model is in no way required to be a physics model. For an example of an implementation of a non-physics model, see Appendix \ref{ch:strassen}.\\
\begin{figure}[t]
    \centering
    \includegraphics[width=0.7\linewidth]{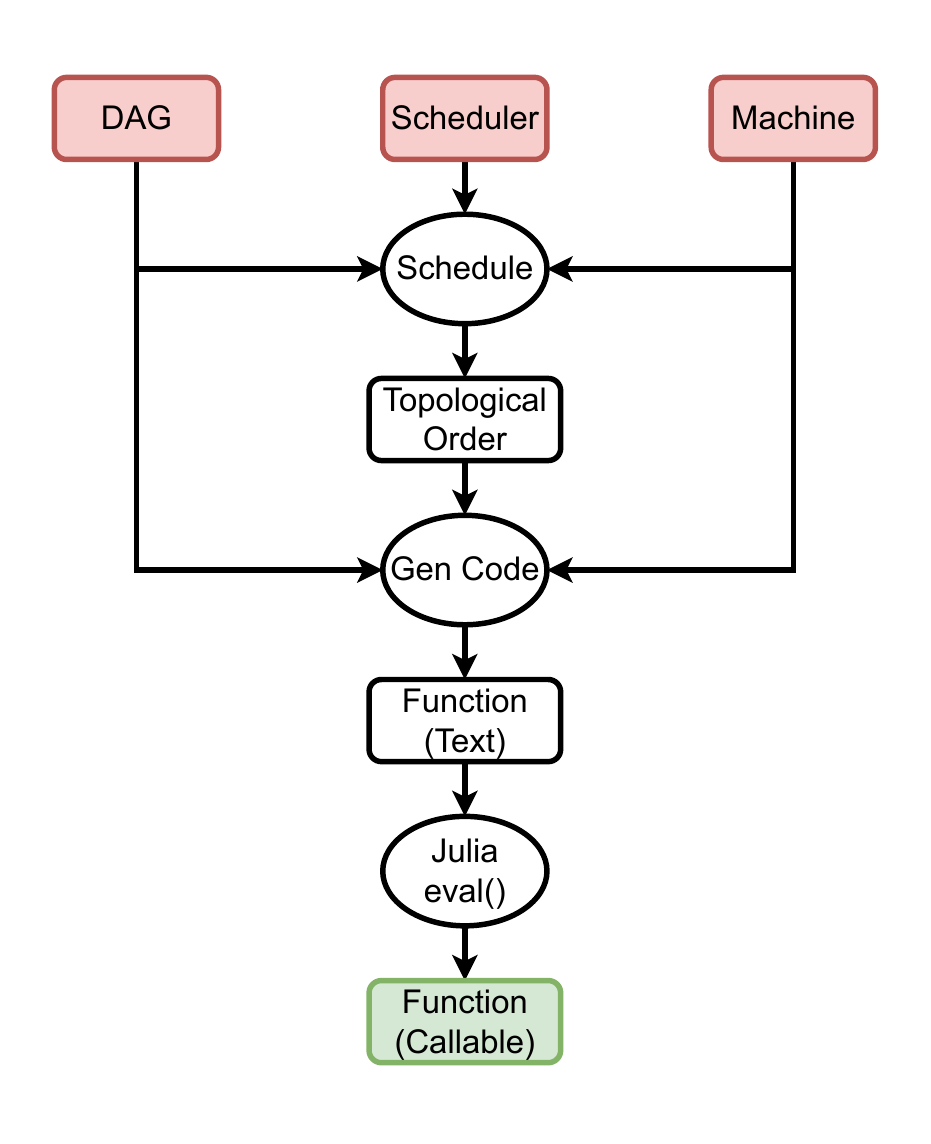}
    \caption{A flowchart showing how a DAG is compiled into a callable function using \texttt{ComputableDAGs.jl}.}
    \label{fig:callstack}
\end{figure}
Once the generator generates a \textit{CDAG} representing the necessary computation, \textit{estimators} can predict the cost to execute the CDAG in its current state. This estimator can be very simple and fast, for example, by simply adding the compute efforts of all tasks, or very complex, for example, by taking the scheduler and devices used into account and running benchmarks.\\
\textit{Optimizers} can then use those estimations to lower the CDAG's cost, using the node \textit{operations}.\\
Once the CDAG has been sufficiently optimized, the \textit{scheduler} can use information about available hardware \textit{devices} to create a schedule. This schedule is a topological ordering of the graph together with a mapping of each compute node to a device.  Note that this means that we only schedule CDAGs statically, i.e., before any computation is executed. This removes the need for a scheduler that is highly optimized for workload distribution during runtime, which is itself a complex problem.\\
From this schedule, the \textit{code generator} generates code to execute the entire CDAG's computation. That \textit{code} then executes on the available devices.\\
\subsection{1-Photon Compton Example}\label{sec:compt_example}
To create a CDAG, we need to translate the Feynman diagrams into the computation they represent. We can do this by replacing each part of each diagram with a respective kernel. The Feynman rules for QED define these kernels; they consist mostly of matrix multiplications. The resulting CDAG takes the four-momenta of all particles participating in the scattering process and computes a single real value.\\
\begin{figure}[tp]
    \centering
    \includegraphics[width=0.68\linewidth]{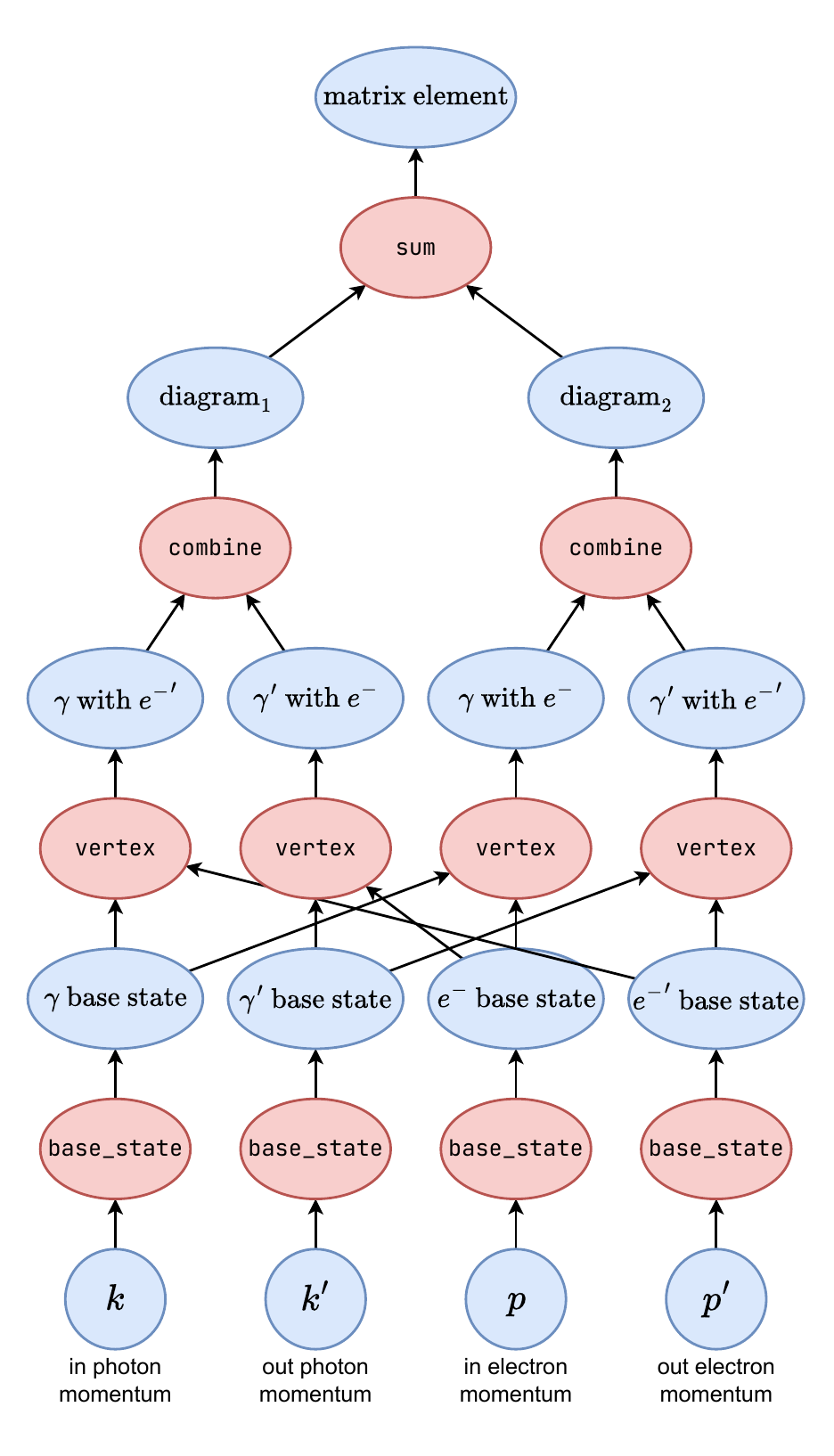}
    \caption{The CDAG to compute the matrix element of the scattering process $e^- \gamma \to {e^-}' \gamma'$ at tree-level for a single spin and polarization combination. Nodes in blue are data nodes and nodes in red are compute nodes. The entry nodes (bottom row) contain the momenta for each of the particles participating in the process. Next, each \texttt{base\_state} is calculated. Then, the four possible combinations of electron-photon combinations are made. From these, two full diagram values are calculated and finally added to yield the matrix element.}
    \label{fig:ke_ke_cdag}
\end{figure}
To convert a Feynman diagram into a CDAG, the following compute kernels are necessary: \begin{itemize}
    \item \texttt{base\_state} functions for every particle species and respective spin/polarization combination,
    \item \texttt{vertex} functions, multiplying two subdiagrams together with a vertex term provided by the QFT and type of vertex,
    \item \texttt{propagator} functions, propagating an inner line of a Feynman diagram,
    \item \texttt{join} functions, joining two halves of a diagram by propagating one side only and multiplying the two sides, and
    \item a \texttt{sum} function, adding all contributing diagram or subdiagram values.
\end{itemize}
For the example process $e^- \gamma \to e^- \gamma$, for which the two tree-level Feynman diagrams shown in Figure \ref{fig:ex_compton} exist, the resulting CDAG is given in Figure \ref{fig:ke_ke_cdag}. The resulting code generated by \texttt{ComputableDAGs.jl} for this process can be found and is explained in Appendix \ref{ch:gen_code_ex}. For the code, we use abbreviations of the task types as follows: \begin{itemize}
    \item \texttt{base\_state} $\to$ U
    \item \texttt{vertex} $\to$ V
    \item \texttt{propagator} $\to$ S1
    \item \texttt{join} $\to$ S2
    \item \texttt{sum} $\to$ Sum
\end{itemize}

\section{Results}\label{ch:results}
The machine used for banchmarking was equipped with an AMD EPYC 7452 32-Core Processor and $\qty{258}{\giga\byte}$ of RAM. For GPU tests, an NVIDIA A30 with \qty{24}{\giga\byte} was used.\\
All of the results were measured using the Julia package \texttt{BenchmarkTools.jl} \cite{BenchmarkTools}, with results collected over a span of  $\qty{120}{\second}$ total for each data point. Julia was invoked with four threads, but the benchmarked functions do not employ any multithreading. For the GPU tests, CUDA libraries at version 12.4.0 were used. Some raw data can be found in Table \ref{tab:raw_data}.\\
\begin{table}[tbp]
\centering
\caption{CDAG sizes, function generation times, and execution times for different n-photon Compton scattering processes in our QED implementation.}
\begin{tabular}{cccc}
process                       & node count & \begin{tabular}[c]{@{}c@{}}median function\\generation time\end{tabular}  & \begin{tabular}[c]{@{}c@{}}median \\execution time\end{tabular} \\
$e^- \gamma \to e^- \gamma$   & 26         & $\qty{8.40}{\milli\second}$                                               & $\qty{427}{\nano\second}$                                       \\
$e^- \gamma^2 \to e^- \gamma$ & 77         & $\qty{11.0}{\milli\second}$                                               & $\qty{203}{\micro\second}$                                      \\
$e^- \gamma^3 \to e^- \gamma$ & 356        & $\qty{17.0}{\milli\second}$                                               & $\qty{20.2}{\micro\second}$                                     \\
$e^- \gamma^4 \to e^- \gamma$ & 2183       & $\qty{47.0}{\milli\second}$                                               & $\qty{88.1}{\micro\second}$                                     \\
$e^- \gamma^5 \to e^- \gamma$ & 15866      & $\qty{286}{\milli\second}$                                                & $\qty{510}{\micro\second}$                                      \\
$e^- \gamma^6 \to e^- \gamma$ & 131069     & $\qty{2.51}{\second}$                                                     & -                                                               \\
$e^- \gamma^7 \to e^- \gamma$ & 1209632    & $\qty{31.5}{\second}$                                                     & -                                                               \\
$e^- \gamma^8 \to e^- \gamma$ & 12337955   & $\qty{895}{\second}$                                                      & -                                                               
\end{tabular}
\label{tab:raw_data}
\end{table}
For the following practical results, we consider the tree-level processes of QED for $e^- \gamma^n \to e^- \gamma$ processes, i.e., n-photon Compton scattering processes.
\subsection{Complexity Analysis}
\begin{figure}[tp]
    \centering
    \includegraphics[width=0.85\linewidth]{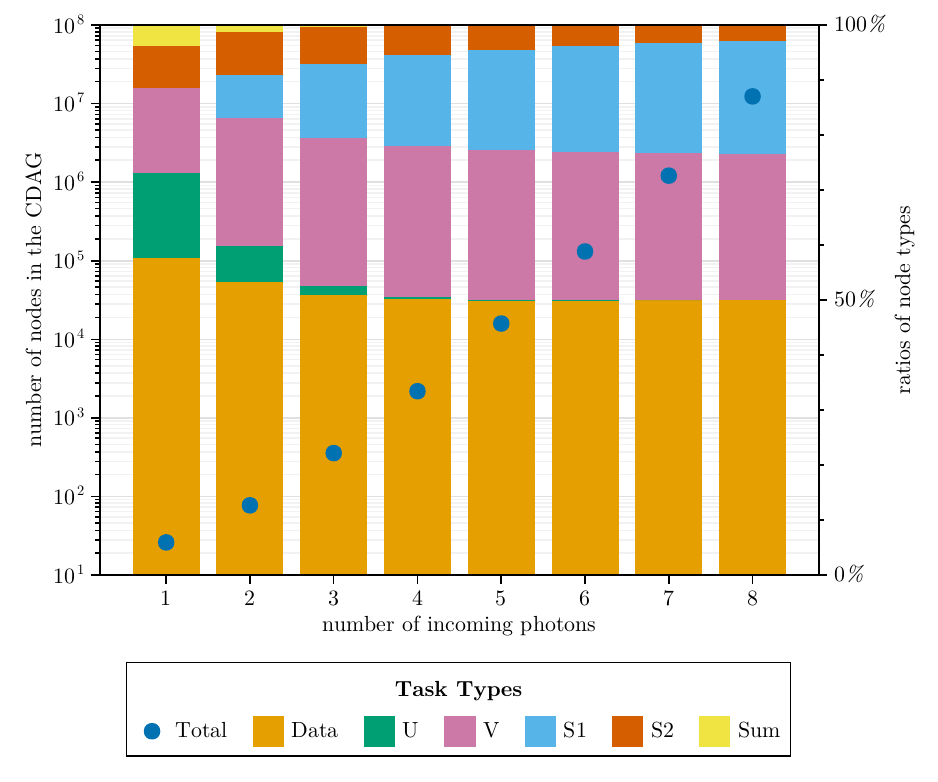}
    \caption{The left axis and blue dots show the number of nodes in a generated CDAG for the matrix element calculation of a $e^- \gamma^n \to e^- \gamma$ QED scattering process. The slight upwards curvature of the plot in log-scale shows the super-exponential scaling. The stacked bars show the ratio of the different task types that make up the CDAG. For an explanation of the individual task types, refer to Section \ref{sec:compt_example}.}
    \label{fig:graph_size}
\end{figure}

As a simple case of QED scattering processes, in an n-photon Compton scattering process (i.e., $n$ incoming photons), $(n+1)!$ contributing Feynman diagrams exist, representing all orderings in which the $n+1$ total photons ($n$ incoming + 1 outgoing) can bind to the electron. We only consider fixed spin and polarization states for all external particles.\\
The algorithm for the DAG generation uses a strategy of reusing reoccurring subdiagrams, yielding CDAG sizes smaller than would naively be expected. However, this technique alone does not lower the computational complexity below a factorial. This is visible in Figure \ref{fig:graph_size}, which shows the size of a generated CDAG for scattering processes $e^- \gamma^n \to e^- \gamma$. Since the y-axis of the graph is in log-scale, the upwards curvature of the slope indicates a super-exponential scaling. To achieve an only exponential scaling, a more sophisticated generation of the DAG or term rewriting systems must be employed. However, for the purpose of testing \texttt{ComputableDAGs.jl} against large problem instances, this scaling is beneficial.\\
The stacked bars in the same figure show the ratio of each task type of the whole graph. For an explanation of the task types, refer to Section \ref{sec:qed}.\\
The biggest part of nodes is made up of data tasks. This is because every data node can only have at most one predecessor, which must be a compute node. The only data nodes that can have no predecessors are the CDAGs entry nodes. These additional entry nodes explain the initial overhead of data nodes for very small graphs, before the ratio tends towards $\qty{50}{\percent}$ of the nodes for larger CDAGs.\\
U tasks are the \texttt{base\_state} tasks, which are generated once per external particle. That means the number of U tasks only scales linearly in comparison to the super-exponential scaling of the total number of nodes. Therefore, the ratio quickly approaches $\qty{0}{\percent}$. The Sum tasks behave similarly, since there is only a single Sum task per CDAG.\\
The remaining task types V, S1, and S2 all scale approximately the same. There is approximately one S1 or S2 task per V task, so their ratio stabilizes at approximately $\qty{25}{\percent}$ V tasks and $\qty{25}{\percent}$ S1 and S2 tasks combined.\\
\subsection{Function Setup Time}
\begin{figure}[t]
    \centering
    \includegraphics[width=0.85\linewidth]{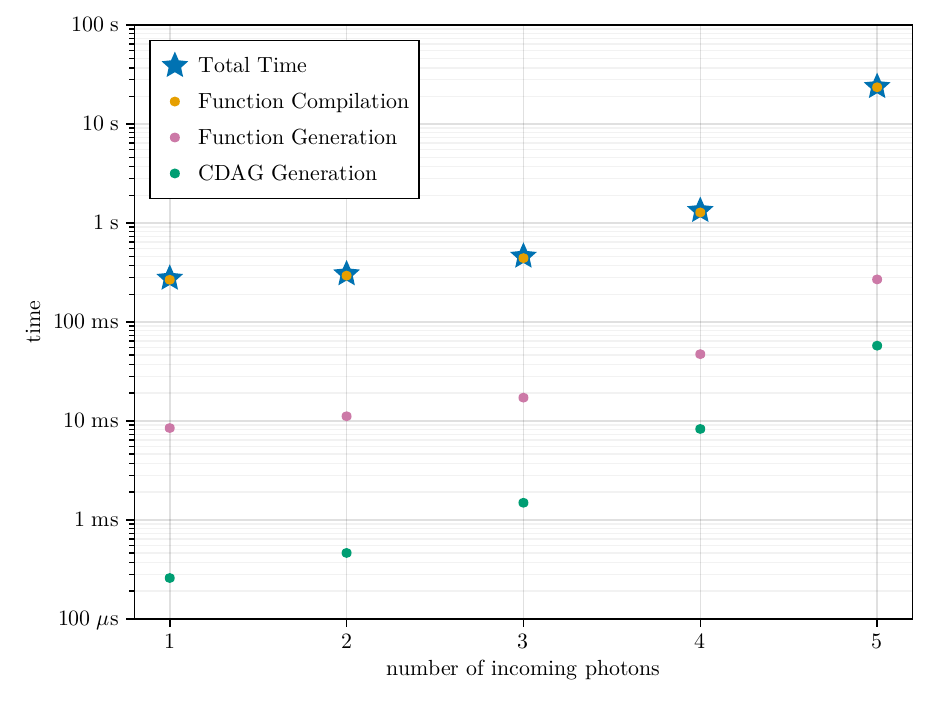}
    \caption{The median times taken to generate a CDAG, generate a function from the CDAG, compile the function, and the total time for a $e^- \gamma^n \to e^- \gamma$ QED scattering process.}
    \label{fig:f_gen}
\end{figure}

Figure \ref{fig:f_gen} shows the time taken to generate a Julia function for a scattering process. Shown are the CDAG generation time, i.e., running the code initially assembling the graph, the function generation time, i.e., translating the CDAG into Julia code, the function compilation time, and the sum of these three stages. Despite the logarithmic y-axis scale, the slopes curve upwards, again indicating the super-exponential scaling. The total time is dominated by the compile time for all process sizes, followed by the function generation time, and finally the CDAG generation time. The function compilation grows especially quickly because compilation of a code block typically takes more than a linear amount of time in the number of lines of code, which already scales super-exponentially.\\
This means that the most important focus for being able to execute and scale to larger CDAGs should be improving the compilation of the generated functions. Some ideas and progress towards this is discussed in the conclusion below.\\
\subsection{Execution Time}\label{sec:exec_results}
\begin{figure}[t]
    \centering
    \includegraphics[width=0.85\linewidth]{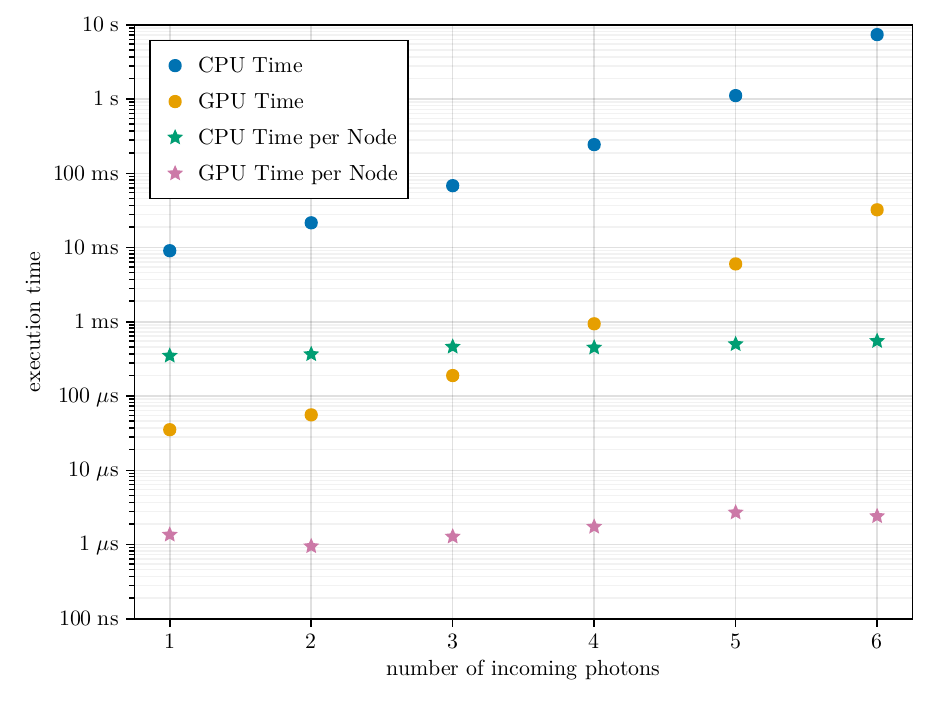}
    \caption{The median times of CPU and GPU execution for $2^{14} = 16384$ calculated matrix elements for $e^- \gamma^n \to e^- \gamma$ scattering processes in QED.}
    \label{fig:f_exec_cpu_vs_gpu}
\end{figure}

In Figure \ref{fig:f_exec_cpu_vs_gpu}, the execution times for the same $e^- \gamma^n \to e^- \gamma$ scattering processes are depicted for CPU and GPU execution, and average computation times per node. The scaling of the total execution time follows that of the number of nodes, as would be expected since the types of nodes stay the same and the ratio of node types converges quickly for scattering processes of increasing size. Additionally, the same plot shows the average execution times per node. The time per node stays relatively stable across the different tested process sizes.\\
The GPU execution uses \texttt{ComputableDAGs.jl}'s ability to compile a CDAG into a kernel to be executed on the GPU. In this experiment, we use an NVIDIA A30 GPU. Note that these results use an optimized version of the generated CDAGs by applying all possible node reductions, before creating a function or GPU kernel.\\
Comparing the GPU results to the CPU results, for the smaller scattering processes, the execution on the GPU is between \num{250} and \num{400} times faster than the CPU execution. For the larger processes, the difference seems to converge to a factor of around \num{200}. Even though it is hard to theoretically predict the expected performance in a comparison between a CPU and a GPU, going purely by theoretical peak FLOPS number \footnote{Note that we use FLOPS for floating point operations per second and FLOPs for floating point operations.} reported in hardware data sheets, this difference in speed seems plausible and indicates that no major losses of performance are introduced by Julia or \texttt{CUDA.jl} when compiling for GPU targets.\footnote{While there are many challenges when comparing CPU to GPU performances, going by rough theoretical numbers, we can find a reported theoretical peak performance of \qty{5161}{\giga\text{FLOPS}} (double precision) for the A30 \cite{a30specs}, and a single thread of an AMD EPYC\texttrademark \space 7452 processor has a theoretical peak performance of about \qty{26.8}{\giga\text{FLOPS}}, assuming full AVX usage doing FMA operations at boost clock speed \cite{amd_specs}. Therefore, we expect a factor of about \num{193} between the GPU and CPU execution speed. This matches the observed factor of \num{200} fairly closely.} The experiment also shows that, while it is generally inadvisable \footnote{Very large kernels start to take up a lot of memory and, depending on the structure of the compiled function, tend to use a lot of local memory through register spilling. This can slow down the execution and lower GPU occupancy.}, very large kernels can be generated and executed on the GPU. However, at a certain point, limits still exist for the size of a GPU kernel, such as a limit of \qty{512}{\kibi\byte} of spilled register storage per thread, or \num{2000000} instructions per kernel in the CUDA version we used.\\
\subsection{CDAG Optimization}
\begin{figure}[tp]
    \centering
    \includegraphics[width=0.85\linewidth]{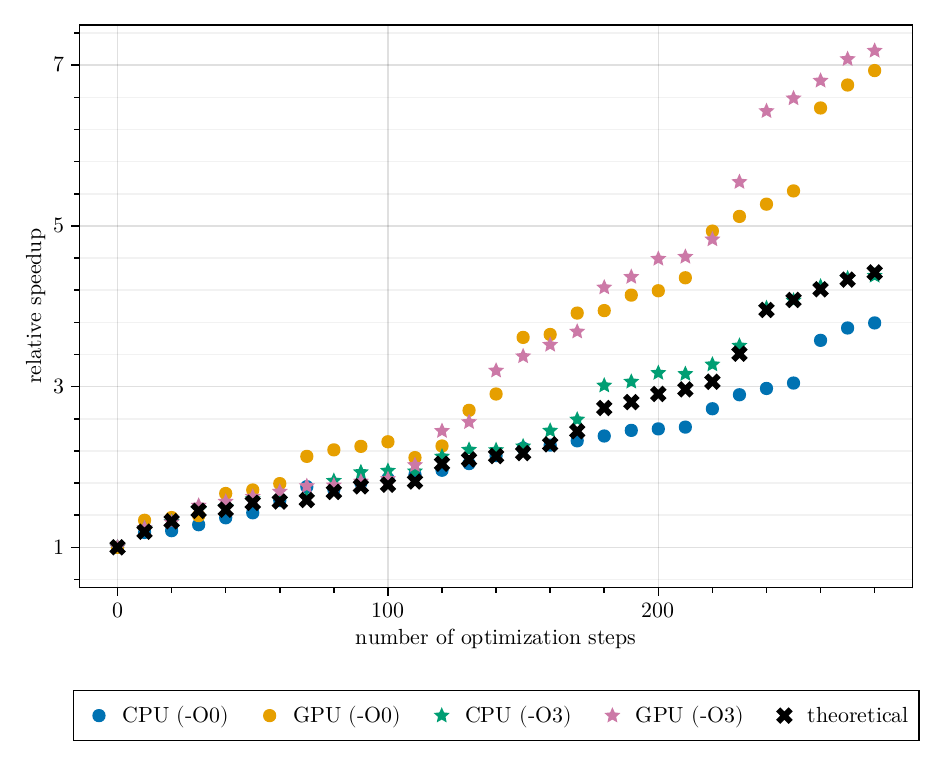}
    \caption{CPU and GPU speedup of a $e^- \gamma^4 \to e^- \gamma$ QED scattering process. The x-axis shows the number of performed optimization steps; the y-axis shows the obtained execution speedup normalized to the initial, i.e., unoptimized state of the CDAG. The black crosses show the speedup calculated from the number of theoretical FLOPs required to calculate the result, which, in turn, is obtained by summing over each compute node's individually measured FLOP count. This theoretical calculation ignores many real world effects, such as data movements, caching or instruction-level parallelism. In this case, this means that especially the GPU heavily profits from the optimizations in other ways (reduced code size, fewer register spills), leading to much better speedup than expected. More sophisticated estimation techniques could incorporate more information to make better predictions.}
    \label{fig:f_optimization}
\end{figure}

In this section, we investigate the effectiveness of optimizations on the CDAG-level, specifically of the node reduction operation. In short, a node reduction merges several nodes that can be deduced to have the same result into one. Figure \ref{fig:f_optimization} shows measured and theoretical performances, each relative to their unoptimized state. The shown measures are CPU performance, GPU performance, each with compiler flags \texttt{-O0} or \texttt{-O3} set, and the theoretical speedup calculated from the FLOPs of the CDAG, summing each compute node's measured FLOPs together \footnote{The individual functions' FLOPs were measured using LIKWID \cite{likwid} and the respective Julia package \texttt{LIKWID.jl}\cite{likwidjl}.}.\\
In the graph's x-axis, the number of applied optimizations is shown. Each optimization here represents a single node reduction, but a node reduction always reduces the maximum possible number of nodes together in one step. The amounts of nodes that can be reduced per step and their types lead to different changes in the resulting performance increase in FLOPs or measured performance. Both the \texttt{-O3} CPU and GPU execution times follow the same curve as the predicted performance from FLOPs with some deviation towards better-than-predicted performance. For the CPU, the fully optimized graph almost perfectly matches the expected performance predicted by the FLOPs reduction, which indicates that in half-optimized states, the compiler can find some optimizations that will later be applied by \texttt{ComputableDAGs.jl} too.\\
In the case of the \texttt{-O3} GPU, the performance starts to show larger differences from what the FLOPs would predict. This is due to the nature of GPUs, which profit more heavily from reduced cache sizes and fewer spilled registers. As already mentioned in Section \ref{sec:exec_results}, the generated kernels are very large and, therefore, many registers get spilled into local memory which physically resides in the main device memory and uses the same caches.\\
It is important to note here that we have deliberately turned off the \texttt{always\_inline} CUDA.jl option. This flag forces the compiler to inline every single function call and produce a completely flat device code. This tends to lead to larger kernels but much faster code, at least for this specific code, with an approximate $\num{11}\times$ speedup compared to the fully optimized CDAG compiled without the flag. Further investigating the compiled device code shows that this flag allows the compiler to remove any duplicated instructions (i.e., essentially perform node reduction on instruction level), which leads to the same device code produced, regardless of the optimization state of the CDAG. This makes it a very powerful option, however, it prevents us from profiling the effects of the CDAG's optimization state on different platforms, which is why we turned it off for this experiment. Critically, this does not mean that node reductions are superfluous, because first, inlining all code can lead to much larger device code, which may not always be desirable, second, because this does not work for heterogeneous computation on multiple threads and external accelerator devices, and third, because not all targets currently support the flag, such as AMDGPU.jl, the Julia library for programming GPUs from AMD.\\
Figure~\ref{fig:f_optimization} also shows the results using the \texttt{-O0} compiler flag. This leads to the CPU performance curve following the FLOPs curve even more closely, with a small linear offset. The final execution is slightly slower than predicted, which is explained by the missing optimizations. The CUDA device compiler seems to be barely affected by the compiler flag. Note that the order of applied optimizations is not deterministic and can differ between the compiler flags, which explains the slightly different course in the middle of the curves between the two. The fully optimized CDAG is, however, still identical between the two, since it is the fixpoint state of the reduction algorithm and does not depend on the order of prior optimizations.\\
\subsection{Breaking Even}
\begin{figure}[t]
    \centering
    \includegraphics[width=0.85\linewidth]{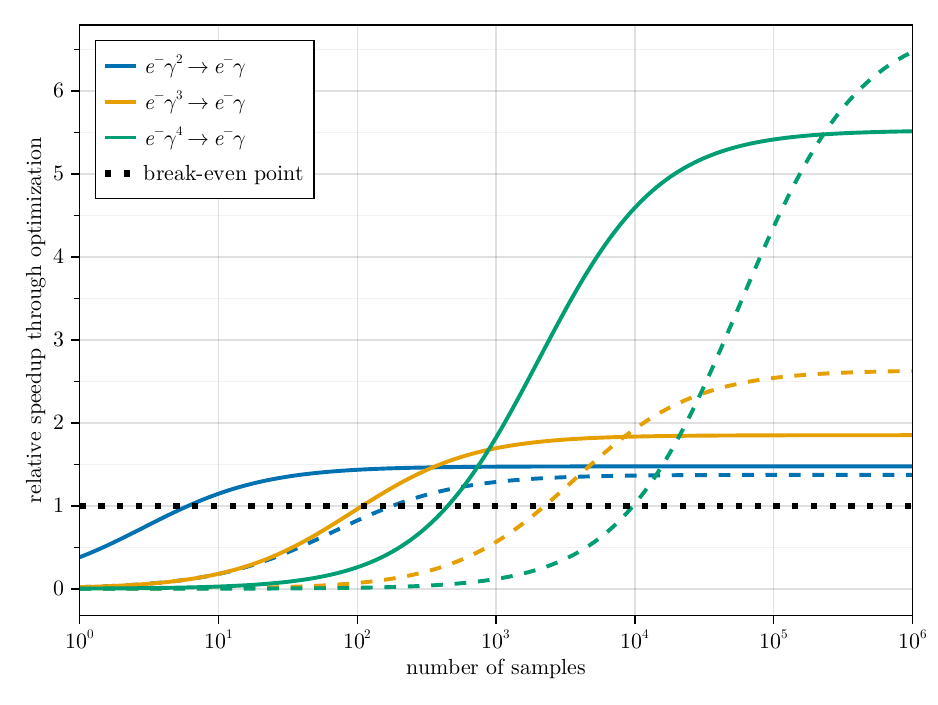}
    \caption{CPU (line) and GPU (dashed) speedup ratios when calculating $N$ samples, including the time taken to optimize. The CPU was benchmarked for single elements, while the GPU used \qty{16384} element batches. We linearly extrapolated the runtime for $N$ elements from the single element runtimes. The black dotted line shows the break-even point; above this line, the optimization amortizes the time it took.}
    \label{fig:f_speedup}
\end{figure}

Figure \ref{fig:f_speedup} shows the speedup from optimization using node reductions, including the time taken to optimize. For a given scattering process, we have the evaluation time $t_e$ for the evaluation of a single element, and $t'_e$ for the evaluation of a single element using the optimized method. The total time to calculate $N$ samples is then $T_e = t_e * N$ for the unoptimized case, and $T'_e = t'_e * N + t_o$, where $t_o$ is the time taken to optimize. The speedup can then be calculated using $S = {T_e / T'_e} = {t_e * N / (t'_e * N + t_o)}$, which approaches ${t_e / t'_e}$ for sufficiently large $N$. This can be seen in the graph, as the curves approach a constant number as $N$ grows.\\
The break-even point in the figure shows the point at which the speedup is $1$, i.e., the point at which $T_e = T'_e$. Where this point lies is dependent on the absolute time saved compared to the time taken to optimize, specifically it is reached at $N_{\mathrm{break-even}} = {t_o / (t_e - t'_e)}$. This is in contrast to the speedup itself, which is dependent on the quotient between $t_e$ and $t'_e$.
For smaller processes like the $e^- \gamma^2 \to e^- \gamma$, the speedup is relatively small. However, since the CDAG for it is also small, the optimization does not take much time either. Therefore, the break-even point is reached very quickly for very small numbers of samples, \qty{6} for the CPU and \qty{179} on the GPU. The larger $e^- \gamma^3 \to e^- \gamma$ process reaches speedups of close to $2\times$ on the CPU and over $2.5\times$ on the GPU. Finally, for $e^- \gamma^4 \to e^- \gamma$, the speedups are about $5.5\times$ on the CPU and close to $7\times$ on the GPU. For these larger processes and therefore larger CDAGs, the optimization starts to take significant amounts of time, leading to the break-even points and a steadying speedup being reached only for larger $N$.\\
Notably, these plots look very similar to the plots of speedups achieved through the parallelization of parts of otherwise serial code, known as Amdahl's law \cite{amdahl}. The formula for the speedup is given by $S = t_{\text{serial}} / (t_{\text{serial part}} +  t_{\text{parallel part}} / N_{\text{processors}})$, where $t_{\text{serial}}$ is the time to execute everything serially, $t_{\text{serial part}}$ is the time to serially execute non-parallelizable parts of the code, and $t_{\text{parallel part}}$ is the time to serially execute parallelizable parts of the code, which are assumed to be perfectly scalable to $N_{\text{processors}}$ processors. Plotting this formula over the number of processors gives similar curves as our plot for the number of samples, since it also rises quickly (linearly) for small numbers of processors, while it tends toward a constant maximal limit for many processors.\\
These observations reinforce the point that the usefulness of our optimizations and the framework of CDAGs in general depends on a use-case needing many thousands, millions or even billions of executions of the resulting function. For matrix element generation in particle physics, this is the case, as the results are commonly used for Monte-Carlo sampling algorithm, requiring very large numbers of samples.

\section{Discussion}\label{ch:discussion}
The experimental results strongly suggest that the generation of CDAGs is a viable approach for representing complex computations in a way that allows optimizations to be done using domain knowledge, while the project and its interfaces itself remain domain-agnostic. Large speedups could be observed using only node reductions, and portability between different architectural targets is seamless, thanks to Julia's first class support of GPUs.\\
\texttt{ComputableDAGs.jl} itself is completely domain-agnostic, as shown by our implementation of two particle physics applications and a linear algebra application. This opens the way for many more implementations of real-world applicable particle physics problems, but also in other fields of science, where, for example, simulations or data processing can often be represented as DAG structures as well. These can then profit from the same logic put into \texttt{ComputableDAGs.jl}, including future improvements and additions thanks to the orthogonal interfaces, without having to handle it themselves.\\
Julia proved versatile and powerful as a development language for \linebreak\texttt{ComputableDAGs.jl}, supported by its metaprogramming capabilities, multiple dispatch for interface design, GPU support, and code analysis tooling, such as \texttt{Cthulhu.jl}, \texttt{BenchmarkTools.jl}, \texttt{PProf.jl}, \texttt{Revise.jl}, among many more. In addition, the versatility of Julia allows us to combine the zero runtime overhead of static scheduling with many of the benefits of dynamic scheduling, such as testing for otherwise hard-to-predict effects of certain schedules, by simply running different schedules, evaluating and comparing them against each other. The same is true for optimizations on the DAG level, whose effects can be hard to predict, due to changes in both data transport and compute.\\
Future work may focus on adding more sophisticated implementations to the interfaces. More sophisticated estimators could integrate more information from the compute nodes, such as expected register pressure, required intermediate memory or expected stack size and depth. Additionally, heterogeneous scheduling should be added to allow the use of different devices for parts of a CDAG. This could allow large speedups by automatically offloading workloads to accelerators that are well suited for them, while continuing to do long serial work on the CPU. With heterogeneous scheduling, estimations can also include information from the data nodes into the expected performance, informing optimizers more accurately on the effects of certain operations.\\
More types of node operations are conceivable. A node vectorization could merge many compute nodes of the same type but with different inputs into one node, applying the function on an array of inputs. Such a vectorized node could lend itself very well for scheduling on a GPU, but even on a CPU it could enable the use of SIMD instructions. Another operation that could be added is term rewriting, which can use properties of nodes such as commutativity, associativity or distributivity to restructure parts of the CDAG. This could itself improve performance or enable other optimizations to follow.\\
More work is also needed to be able to compute even larger CDAGs. A promising path seems to be splitting the generated function itself into multiple smaller functions since compilers struggle with very large serial functions. Enabling heterogeneity and offloading parts of a CDAG to other devices will also likely help with the compile time problems.\\
Targeting large multi-node HPC systems will be part of future research as well. This could be achieved by partitioning the CDAG into loosely connected subgraphs of similar size that can then be individually optimized on the different nodes.

\section{Conclusion}\label{ch:conclusion}
In this paper we laid theoretical foundations for a framework of representing large computations as DAGs and showed possible ways to optimize these DAGs on a high level of abstraction and predict their performance. Furthermore, we presented the implementation work done on this theory in form of the Julia package \texttt{ComputableDAGs.jl}, providing the proof of concept of this framework with the generation of CDAGs for a relevant problem of particle physics, their optimization and analysis, and execution on different target architectures. With the established interfaces, other optimization targets such as memory footprint or energy consumption can be integrated in the future.

\section*{Acknowledgment}
This work was partly funded by the Center for Advanced Systems Understanding (CASUS) that is financed by Germany’s Federal Ministry of Research, Technology and Space (BMFTR) and by the Saxon Ministry for Science, Culture and Tourism (SMWK) with tax funds on the basis of the budget approved by the Saxon State Parliament.\\
Additional thanks goes to Klaus Steiniger for proofreading the draft of this paper and giving valuable feedback.\\

\appendix

\section{Conventions}\label{ch:conventions}
Here, we briefly list the mathematical notations and conventions used:

\begin{itemize}
    \item $\mathbb{N}$ denotes the positive natural numbers, including $0$.
    \item $\overline{\mathbb{R}} := \mathbb{R} \cup \{-\infty, +\infty\}$ denotes the extension of the real numbers with positive and negative infinity.
    \item Subscripts on a tuple variable reference the $n$-th element of the tuple, starting at $1$.
    \item Set operations have no precedence over one another; they are applied in order from left to right.
    \item $\mathcal{P}(X)$ denotes the power set of $X$.
    \item Any DAG we use is finite.
\end{itemize}

\section{Formal Background}\label{ch:formal_background}

\subsection{Tasks}
In the present work, a task $T$ is an abstract representation of a particular piece of work for the computer that must be completed. We denote the set of all tasks $\mathbb{T}$ and assume that it is not empty: $\mathbb{T} \neq \emptyset$. All tasks used in this work are denoted $T$ and in $\mathbb{T}$.\\
We consider two fundamentally different types of tasks: \textit{Data Tasks} and \textit{Compute Tasks}. A data task represents a data transport from one computing device to another without muting the data. A compute task represents any calculable compute function. The input and output of those tasks will be discussed below in the context of graphs.\\
We define the set of data tasks $\mathbb{T}_D$ and the set of compute tasks $\mathbb{T}_C$ to have the following properties:
\begin{enumerate}
    \item Both sets are true subsets of the set of all tasks: $\mathbb{T}_D \subsetneq \mathbb{T}$ and $\mathbb{T}_C \subsetneq \mathbb{T}$.
    \item Both sets are not empty: $\mathbb{T}_D \neq \emptyset$ and $\mathbb{T}_C \neq \emptyset$.
    \item The union of both sets makes up all tasks: $\mathbb{T}_D \cup \mathbb{T}_C = \mathbb{T}$.
    \item The intersection of both sets is empty: $\mathbb{T}_D \cap \mathbb{T}_C = \emptyset$.
\end{enumerate}
Combined, these properties also mean that $\{\mathbb{T}_D, \mathbb{T}_C\}$ is a partition of $\mathbb{T}$.\\
Next, we define a \textit{Task Effort} function $\epsilon: \mathbb{T} \rightarrow \mathbb{N}$ as a measure of the size of the task. Since the size of a data task is not comparable to the size of a compute task, we define function restrictions for $\epsilon$ on $\mathbb{T}_D$ and $\mathbb{T}_C$:
\begin{itemize}
    \item The \textit{Data Size} $d:= \restr{\epsilon}{\mathbb{T}_D}$ gives the number of Bytes a data task must transfer.
    \item The \textit{Compute Effort} $c:= \restr{\epsilon}{\mathbb{T}_C}$ gives the number of floating point operations necessary to compute the task.
\end{itemize}
Both the data size and the compute effort of a given task can vary depending on various factors, such as the hardware architecture and the programming language used. For now, we assume these values are given and constant for a task.\\
We define two special identity tasks $\text{ID}_D \in \mathbb{T}_D$ and $\text{ID}_C \in \mathbb{T}_C$ with $e(\text{ID}_D) = e(\text{ID}_C) = 0$. These identity tasks are unique and, therefore, the respective kernels of $\mathbb{T}_D$ and $\mathbb{T}_C$. Their mandatory existence also guarantees that the sets are not empty.\\

\subsection{Directed Acyclic Graph}
We define a \textit{Directed Graph}, or short \textit{DG}, as follows: $\text{DG} = (V, E)$, where $V \subseteq \mathbb{T}$ are the nodes of the graph, with $V_D := V \cap \mathbb{T}_D$ and $V_C := V \cap \mathbb{T}_C$. $E \subseteq (V_D \times V_C) \cup (V_C \times V_D)$ is the set of the graph's directed edges, only allowing edges from compute to data tasks or vice versa.\\
A node $v \in V$ is called an \textit{Entry Node} if and only if no edge $e \in E$ with $e = (v_1, v)$ exists, i.e., no edge in the graph points into $v$. Similarly, a node is called \textit{Exit Node} if and only if no edge $e \in E$ with $e = (v, v_2)$ exists, i.e., no edge in the graph points out of $v$. A DG must have at least one entry node, and for simplicity, we assume that a DG has exactly one exit node \footnote{
    It can be useful to allow multiple exit nodes for simulations where large parts of the calculation can be reused across different simulations with individual results. However, it is always possible to batch up multiple data nodes, resulting in a single exit node outputting multiple individual results.
}.\\
In order to define a path, we first define a convenient incidence function $\phi:E\rightarrow V\times V$ with $\phi(e) = \phi((e_1, e_2)) = (e_1, e_2)$ \footnote{We use a subscript $n$ on a tuple to reference the $n$-th element in the tuple. See Appendix \ref{ch:conventions}. While $\phi$ is essentially an identity function, it helps to make notation less ambiguous.}, returning the tuple of an edge's incident nodes. This allows us to define a path $P$ in a graph as a sequence of edges $P \in E^n, P = (e_1, e_2, \dots, e_n)$, where for $i \in \{1, \dots, n-1\}: \phi(e_i)_2 = \phi(e_{i+1})_1$. A cycle $C$ is a path with the added condition $\phi(e_1)_1 = \phi(e_n)_2$, i.e., the first and last node of the path are the same. If a DG contains no such cycle, we call it \textit{Acyclic}, and the graph is a \textit{Directed Acyclic Graph}, or short \textit{DAG}.\\
It can be seen that a non-empty directed graph has to contain an exit node or a cycle. This means that the two conditions that our graphs are acyclic and only contain a single exit node already infer that our graphs are connected. Otherwise, there would be cycles in one of the graph's connected subgraphs or multiple exit nodes.\\
We now define $\mathbb{D}$ as the set of all possible DAGs with the mentioned properties. Since $\mathbb{T} \neq \emptyset$ it follows that $\mathbb{D} \neq \emptyset$.\\
For brevity, for a DAG with or without any subscripts, we assume that $V_n$ is the set of vertices and $E_n$ is the set of edges of a $\text{DAG}_n$.

We define the following convenience functions:
\begin{itemize}
    \item $v_D: \mathbb{D} \rightarrow \mathcal{P}(\mathbb{T})$ as $v_D(\text{DAG}) = V \cap \mathbb{T}_D$, the set of data tasks in a DAG,
    \item $v_C: \mathbb{D} \rightarrow \mathcal{P}(\mathbb{T})$ as $v_C(\text{DAG}) = V \cap \mathbb{T}_C$, the set of compute tasks in a DAG,
\end{itemize}
where $\mathcal{P}(X)$ denotes the power set of $X$.

\subsection{Computation}
Every task has a function $f: \mathbb{R}^n \rightarrow \mathbb{R}^m$ mapping from the task's input data to its result. For compute tasks, these functions will work on the data of all their input data and produce output data.\\
For data tasks, these functions do not change the data, i.e., their function is the identity function; however, they can represent three different operations:
\begin{itemize}
    \item "create" data if they are an entry node of the DAG they are in; in practice, this will be generated data from an event generator or read from disk,
    \item "destroy" data if they are an exit node of the DAG they are in; in practice, this will be data written to the output as the result,
    \item do nothing with the data, representing a move between compute nodes.
\end{itemize}
By recursively inserting the prerequisite tasks' data into the exit node's function parameters, we can describe an entire DAG's computation as a function of all its entry nodes' data. We call this function the \textit{Computation} of the DAG. We denote it $f_T: \mathbb{R}^n \rightarrow \mathbb{R}^m$ for a given task $T \in \mathbb{T}$.\\
\begin{exmp}
    Consider the graph in Figure \ref{fig:ex_cdag}. The graph contains three compute task nodes, shown in red: $T_1$ with the function $f_{T_1}(x) := e^x$, $T_2$ with the function $f_{T_2}(x) := 5x - 2$, and $T_3$ with the function $f_{T_3}(x, y) := y * \sin(x)$. The blue nodes mark the data task nodes: $x$ and $y$ serve as inputs, $x'$ and $y'$ transport data between nodes, and $z$ outputs the result. We can compute the DAG's computation by inserting each previous node's result into the next node's function. From compute node $T_1$, it follows that $x' = e^x$ and from compute node $T_2$, it follows that $y' = 5y-2$. Finally, inserting these intermediate results into $T_3$, we arrive at $z = (5y - 2) * \sin(e^x)$. Since $z$ is the DAG's exit node, $f(x, y) := (5y - 2) * \sin(e^x)$ is this graph's computation.
\end{exmp}
The \textit{Result} of a DAG computation is the result of its exit node. We define two DAGs $\text{DAG}_1, \text{DAG}_2 \in \mathbb{D}$ \textit{Computationally Equivalent}, if and only if their computations $f_1$ and $f_2$ are equivalent, i.e., their domains and codomains are equal, and for each $n \in D: f_1(n) = f_2(n)$, where $D$ is the domain. In this case, we denote $\text{DAG}_1 \Leftrightarrow \text{DAG}_2$. This relation is trivially reflexive, symmetric, and transitive, making it an equivalence relation. Similarly, we define two tasks \textit{Computationally Equivalent}, if and only if their computations are equivalent, and denote $T_1 \Leftrightarrow T_2$. This is an equivalence relation, too.\\
Some functions work on a specific task but need the context of the DAG. For this purpose, we define the set $\mathbb{D}_T \subseteq \mathbb{D} \times T$, $\mathbb{D}_T = \{(D, T) \in \mathbb{D} \times \mathbb{T} | T \in D_1\}$, which is the set of all tuples of DAGs and tasks that are vertices in the DAG.
For convenience, we also define the sets $\mathbb{D}_{T_D}$ and $\mathbb{D}_{T_C}$ as $\mathbb{D}_{T_D} = \{(D, T_D) \in \mathbb{D} \times \mathbb{T}_D | T_D \in v_D(D)\}$ and $\mathbb{D}_{T_C} = \{(D, T_C) \in \mathbb{D} \times \mathbb{T}_C | T_C \in v_C(D)\}$.\\
\begin{proposition}
    $\{\mathbb{D}_{T_D}, \mathbb{D}_{T_C}\}$ is a partition of $\mathbb{D}_T$.
    \begin{proof}
        Two sets $A$ and $B$ partition a third set $C$ when neither $A$ nor $B$ are empty, the intersection of $A$ and $B$ is empty, and the union of $A$ and $B$ is the set $C$.
        \begin{itemize}
            \item $\mathbb{D}_{T_D} \neq \emptyset$ and $\mathbb{D}_{T_C} \neq \emptyset$ follow from $\mathbb{D} \neq \emptyset$ and $\mathbb{T}_D \neq \emptyset$ and $\mathbb{T}_C \neq \emptyset$, respectively.
            \item $\mathbb{D}_{T_D} \cap \mathbb{D}_{T_C} = \emptyset$ follows from $\mathbb{T}_D \cap \mathbb{T}_C = \emptyset$.
            \item $\mathbb{D}_{T_D} \cup \mathbb{D}_{T_C} = \mathbb{D}_T$ follows from $\mathbb{T}_D \cup \mathbb{T}_C = \mathbb{T}$.
        \end{itemize}
        Therefore, $\mathbb{D}_{T_D}$ and $\mathbb{D}_{T_C}$ form a partition of $\mathbb{D}_T$.
    \end{proof}
\end{proposition}
Furthermore, cost functions $t_D: \mathbb{T}_D \rightarrow \mathbb{R}$ for data tasks and $t_C: \mathbb{T}_C \rightarrow \mathbb{R}$ for compute tasks must be chosen. From these, we define the general task cost function $t: \mathbb{T} \rightarrow \mathbb{R}$:
\begin{equation}
    t(T) := \begin{cases}
        t_D(T) \text{, if } T \in \mathbb{T}_D, \\
        t_C(T) \text{, if } T \in \mathbb{T}_C.
    \end{cases}
\end{equation}
These cost functions will later be used as input for optimization algorithms muting the graph.

\subsubsection{Global Functions on DAGs}
For convenience, we define a few more functions to work on a DAG. To get a specific task's neighbors, we define $p: \mathbb{D}_T \rightarrow \mathcal{P}(\mathbb{T})$ as
\begin{equation}
    p(\text{DAG}, T) := \{T_P \in \mathbb{T} | (T_P, T) \in E\},
\end{equation}
to return the set of $T$'s predecessors in the $\text{DAG}$. Predecessors of $T$ are the tasks on which $T$ depends. Similarly, we define $s: \mathbb{D}_T \rightarrow \mathcal{P}(\mathbb{T})$ as
\begin{equation}
    s(\text{DAG}, T) := \{T_S \in \mathbb{T} | (T, T_S) \in E\},
\end{equation}
to return the set of $T$'s successors in the $\text{DAG}$. Successors of $T$ are the tasks that depend on $T$.\\
When $p(T) \neq \emptyset$, i.e., $T$ has at least one predecessor, we know that $T \in \mathbb{T}_D \Leftrightarrow p(T) \in \mathcal{P}(\mathbb{T}_C)$ and $T \in \mathbb{T}_C \Leftrightarrow p(T) \in \mathcal{P}(\mathbb{T}_D)$, or in other words, if and only if $T$ is a compute task, all its predecessors are data tasks, and vice versa. The same is true for $T$'s successors. This follows from our definition of edges which only allows edges between compute and data tasks.\\
Specifically for compute tasks, we define $d_i: \mathbb{T}_C \rightarrow \mathbb{N}$, for $T_C \in \mathbb{T}_C$, i.e., the function returning $T_C$'s amount of input data, as
\begin{equation}
    d_i(T_C) := \sum_{T_{D_i} \in p(T_C)}{d(T_{D_i})}.
\end{equation}
Using this definition, the \textit{Compute Intensity} of a task $i: \mathbb{T}_C \rightarrow \overline{\mathbb{R}}$ is defined as
\begin{equation}
    i(T_C) :=
    \begin{cases}
        c(T_C) / d_i(T_C) \text{, if } d_i(T_C) \neq 0, \\
        +\infty \text{, otherwise}.
    \end{cases}
\end{equation}

\subsubsection{Global Evaluation Functions}
We can also extend data transfer and compute effort functions to an entire graph:
\begin{itemize}
    \item \textbf{Data Transfer}: \begin{equation} \label{eq:graph_data_transfer}
              D(\text{DAG}) := \sum_{T \in V \cap \mathbb{T}_D}{d(T) * \left\lvert s(\text{DAG}, T) \right\rvert },
          \end{equation}
    \item \textbf{Compute Effort}: \begin{equation} \label{eq:graph_compute_effort}
              C(\text{DAG}) := \sum_{T \in V \cap \mathbb{T}_C}{c(T)},
          \end{equation}
    \item \textbf{Compute Intensity}: \begin{equation} \label{eq:graph_compute_intensity}
              I(\text{DAG}) := \frac{C(\text{DAG})}{D(\text{DAG})}.
          \end{equation}
\end{itemize}
Note that for the graph's data transfer in Equation \ref{eq:graph_data_transfer}, we need to multiply the amount of data transferred by the task with the number of the node's successors since the task itself only knows about what data to transfer, not to how many targets.

\section{Node Operations}\label{ch:node_op}
The high-level abstraction and meta-representation of a large and modular computation allows analysis and mutation of the CDAG. For analysis, we can assign a \textit{compute effort} to each compute node and a \textit{data transfer} to each data node. These abstract values represent the expected amount of time a computation or data transport may take and should be consistent relative to each other. Of course, static assignments like this can not accurately capture the complexity of real hardware, but it is enough if it can serve as a rough direction for an optimizer.\\
To optimize, we need operations as options. In this case, operations are functions that transform a CDAG into another CDAG of equivalent computation. This assures that while changing the graph's structure, the result of the CDAG will stay equivalent, while its properties, such as its total compute effort, may change.\\
\begin{figure}[t]
    \centering
    \includegraphics[width=\linewidth]{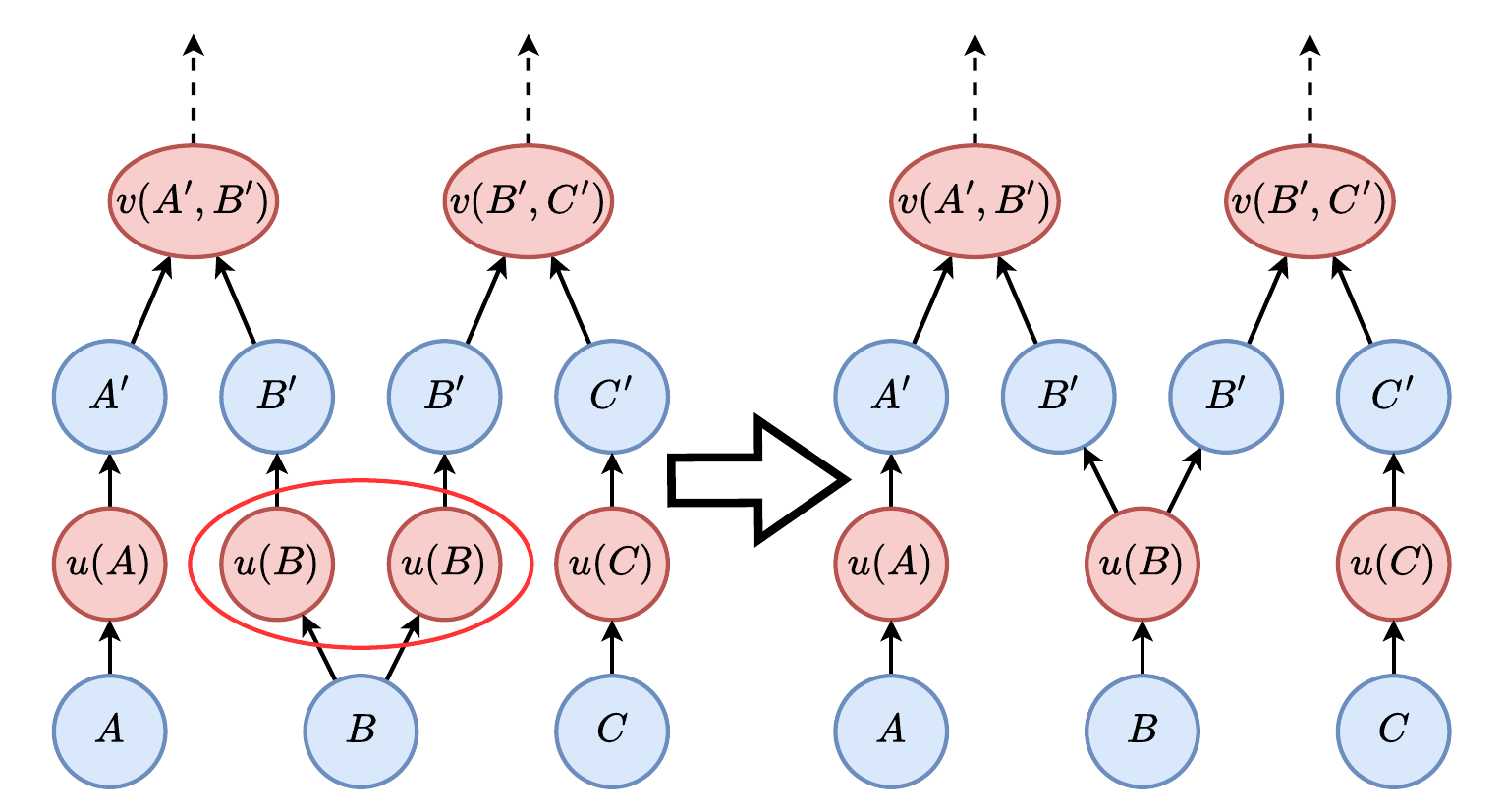}
    \caption{A node reduction. On the left side, two compute nodes have the same parent and compute function; Therefore, they can be reduced into a single compute node with two data node children. Note that this does not contradict the requirement that compute nodes only have one result, because both data nodes receive the exact same data.}
    \label{fig:node_reduction}
\end{figure}
\begin{figure}[t]
    \centering
    \includegraphics[width=\linewidth]{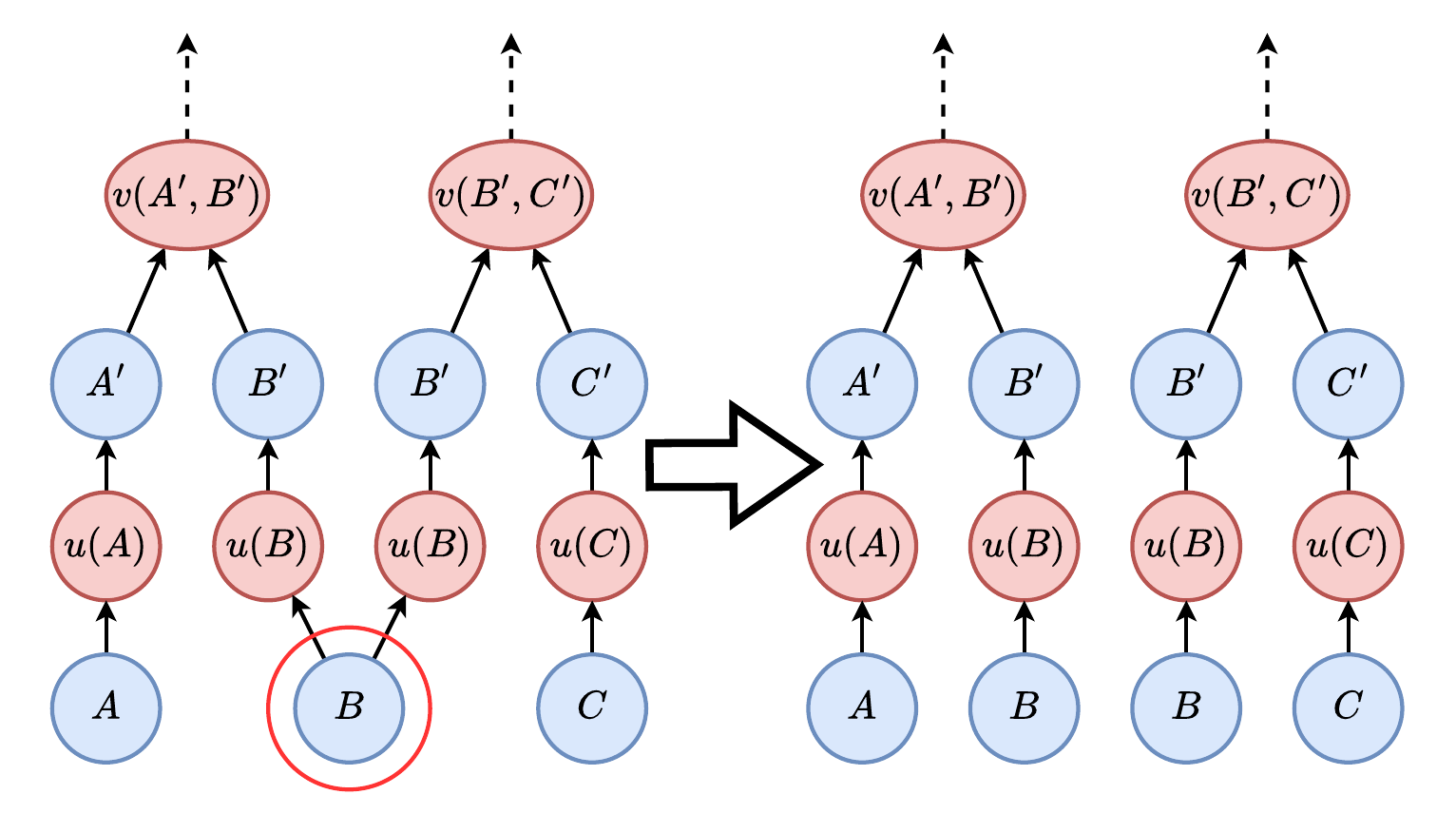}
    \caption{A node split. As the reverse operation of the node reduction shown in Figure \ref{fig:node_reduction}, a node with two children is split into two nodes with equal parents and function, with a single child each.}
    \label{fig:node_split}
\end{figure}
We currently have two such operations implemented: Node split and node reduction.\\
A node reduction is an operation on a CDAG where any number $n$ of nodes are merged into a single node. The requirements for this are that all $n$ nodes have the same set of parent nodes and the same function. If this is the case, all $n$ nodes can be reduced to one node. This new node's children are the union of all children of the reduced nodes. An example is given in Figure \ref{fig:node_reduction}. Since the operation is well-defined, the effects of the operation on the entire CDAG's compute effort and data transfer can easily be calculated. For details, please refer to the relevant section in Appendix \ref{ch:formal_background}.\\
While a node split may seem unintuitive when reducing a CDAG's compute effort or data transfer, it can allow other node operations to be applied afterward and spread computations across more physical devices, saving data transfer effort at the cost of more required computation. Depending on the specific target machine, this can improve the total performance.\\
Additional types of node operations are conceivable: A node vectorization could merge nodes with equal functions but mutually distinct parents. This could allow parallelization on GPUs or SIMD instructions and multithreading on CPUs.\\
If certain attributes of the compute functions assigned to the nodes are given by the developer or inferred by the compiler, using term rewriting mechanisms \cite{blom2001term} as operations would also be possible. For example, associativity, commutativity, and distributivity could be used to restructure the CDAG into a more efficient form or a form that allows more subsequent node reductions or vectorizations.\\

\section{Concrete Example}\label{ch:gen_code_ex}
Here, we show an example of what actual code generated by \linebreak\texttt{ComputableDAGs.jl} looks like. Each generated function is wrapped in a \linebreak\texttt{RuntimeGeneratedFunction} \cite{runtimegeneratedfunctions}:
\begin{lstlisting}[language=Julia]
RuntimeGeneratedFunction(
    #=in ComputableDAGs=#,
    #=using Main=#,
    :((input,)->begin
        # generated code
        return result
    end
)
\end{lstlisting}
The generated code is divided into a section of input assignments, assigning the CDAGs data entry nodes to an expression calculated from the \texttt{input} symbol passed to the function. In this case, we generated a function for the matrix element of a $e^- \gamma \to e^- \gamma$ process, so four entry nodes exist, one for each particle momentum.

\begin{lstlisting}[language=Julia,showstringspaces=false]
ph_out = ($(Expr(:opaque_closure,:(() -> begin
    begin
        local ph_out, var"::", var"ParticleValueSP{ParticleStateful{Outgoing, Photon, SFourMomentum}, PolarizationX, ComplexF64}"
    end
    ph_out = ParticleValueSP(
        ParticleStateful{Outgoing,Photon,SFourMomentum}(momentum(input, Outgoing(), Photon(), Val(1))),
        0.0im,
        PolX(),
    )
    return ph_out
end))))()
\end{lstlisting}
Similar lines follow for the outgoing electron \texttt{el\_out} and the incoming photon and electron \texttt{ph\_in} and \texttt{el\_in} \footnote{Note that the names of the variables have been chosen by us and were inserted here to make the code much easier to understand. The automatically generated variable names are random.}.
Next follows the block of code containing all functions generated by compute and data nodes of the graph.
\begin{lstlisting}[language=Julia, numbers=none, breaklines=false]
el_in_p = (identity)(el_in)
el_in_bs = (ComputableDAGs.compute)(
    ComputeTaskQED_U(), el_in_p
)
el_in_bs_p = (identity)(el_in_bs)
el_out_p = (identity)(el_out)
el_out_bs = (ComputableDAGs.compute)(
    ComputeTaskQED_U(), el_out_p
)
el_out_bs_p = (identity)(el_out_bs)
ph_in_p = (identity)(ph_in)
ph_in_bs = (ComputableDAGs.compute)(
    ComputeTaskQED_U(), ph_in_p
)
ph_in_bs_p = (identity)(ph_in_bs)
ph_in_el_out = (ComputableDAGs.compute)(
    ComputeTaskQED_V(), ph_in_bs_p, el_out_bs_p
)
ph_in_el_out_p = (identity)(ph_in_el_out)
ph_in_el_in = (ComputableDAGs.compute)(
    ComputeTaskQED_V(), ph_in_bs_p, el_in_bs_p
)
ph_in_el_in_p = (identity)(ph_in_el_in)
ph_out_p = (identity)(ph_out)
ph_out_bs = (ComputableDAGs.compute)(
    ComputeTaskQED_U(), ph_out_p
)
ph_out_bs_p = (identity)(ph_out_bs)
ph_out_el_in = (ComputableDAGs.compute)(
    ComputeTaskQED_V(), ph_out_bs_p, el_in_bs_p
)
ph_out_el_in_p = (identity)(ph_out_el_in)
diagram_1 = (ComputableDAGs.compute)(
    ComputeTaskQED_S2(),
    ph_out_el_in_p,
    ph_in_el_out_p,
)
diagram_1_p = (identity)(diagram_1)
ph_out_el_out = (ComputableDAGs.compute)(
    ComputeTaskQED_V(), ph_out_bs_p, el_out_bs_p
)
ph_out_el_out_p = (identity)(ph_out_el_out)
diagram_2 = (ComputableDAGs.compute)(
    ComputeTaskQED_S2(),
    ph_in_el_in_p,
    ph_out_el_out_p,
)
diagram_2_p = (identity)(diagram_2)
diagrams_sum = (ComputableDAGs.compute)(
    ComputeTaskQED_Sum(2),
    diagram_2_p,
    diagram_1_p,
)
diagrams_sum_p = (identity)(diagrams_sum)
\end{lstlisting}
In this code, we can see one assignment generated per node in the CDAG. Every assignment to a variable ending in \texttt{\_p} signifies a data node. In this example, the code is compiled only for CPU, so each data node is merely a renaming of variables. The \texttt{identity} function simply returns its argument which results in no overhead for the runtime of the function. In the case of a data node connecting compute nodes scheduled to different devices, instead of \texttt{identity}, some device specific \texttt{copy} function would be called.
Compute nodes are compiled to calls of the \texttt{ComputableDAGs.jl} interface function \texttt{compute}, together with the type of compute function. The following tasks are defined:
\begin{itemize}
    \item \texttt{ComputeTaskQED\_U}: Compute the base state of a particle from its species, spin or polarization, and momentum,
    \item \texttt{ComputeTaskQED\_V}: Compute the value after combining two subdiagrams by multiplying their states with the vertex term,
    \item \texttt{ComputeTaskQED\_S1}: Compute the propagation of a virtual particle, \footnote{This compute function does not appear in the example, because the diagrams are too small and S2 can be used immediately instead.}
    \item \texttt{ComputeTaskQED\_S2}: Compute the value of combining two subdiagrams that make up a full diagram by multiplying their values and their propagator, and
    \item \texttt{ComputeTaskQED\_Sum}: Compute the sum of all full diagrams' values.
\end{itemize}
In the function we can see all the parts necessary to calculate the matrix element:
\begin{itemize}
    \item The computation of each of the external particles' base states \texttt{ph\_in\_bs}, \texttt{el\_in\_bs}, \texttt{ph\_out\_bs}, and \texttt{el\_out\_bs}.
    \item The computation of the vertices: \begin{itemize}
        \item Combining the incoming photon with the outgoing electron\linebreak\texttt{ph\_in\_el\_out},
        \item combining the incoming photon with the incoming electron\linebreak\texttt{ph\_in\_el\_in},
        \item combining the outgoing photon with the incoming electron\linebreak\texttt{ph\_out\_el\_in}, and
        \item combining the outgoing photon with the outgoing electron\linebreak\texttt{ph\_out\_el\_out}.
    \end{itemize}
    \item The computation of the full diagrams: \begin{itemize}
        \item Combining \texttt{ph\_out\_el\_in} and \texttt{ph\_in\_el\_out} into \texttt{diagram\_1} and
        \item combining \texttt{ph\_in\_el\_in} and \texttt{ph\_out\_el\_out} into \texttt{diagram\_2}.
    \end{itemize}
    \item Finally, the computation of the sum of \texttt{diagram\_1} and \texttt{diagram\_2} into \texttt{diagrams\_sum}, which is the exit node of the CDAG.
\end{itemize}
The final value of \texttt{diagrams\_sum\_p} is then returned from the function.

\section{Strassen Matrix Multiplication Algorithm}\label{ch:strassen}
The Strassen Matrix Multiplication Algorithm is an algorithm for the multiplication of block matrices with a better scaling for very large matrices than the naive algorithm, which scales $\mathcal{O}(n^3)$ \cite{strassen}.\\
Consider the matrices $A, B, C \in \mathbb{R}^{n\times n}$ with $n \in \mathbb{N_+}$ that can be partitioned into equally sized block matrices \[A = \begin{bmatrix} A_{11} & A_{12} \\ A_{21} & A_{22} \end{bmatrix}, B = \begin{bmatrix} B_{11} & B_{12} \\ B_{21} & B_{22} \end{bmatrix}, C = \begin{bmatrix} C_{11} & C_{12} \\ C_{21} & C_{22} \end{bmatrix}. \] To calculate $C = A \times B$, the algorithm defines intermediate values \begin{itemize}
    \item $M_1 = (A_{11} + A_{22}) \times (B_{11} + B_{22})$,
    \item $M_2 = (A_{21} + A_{22}) \times B_{11}$,
    \item $M_3 = A_{11} \times (B_{12} - B_{22})$,
    \item $M_4 = A_{22} \times (B_{21} - B_{11})$,
    \item $M_5 = (A_{11} + A_{12}) \times B_{22}$,
    \item $M_6 = (A_{21} - A_{11}) \times (B_{11} + B_{12})$, and
    \item $M_7 = (A_{12} - A_{22}) \times (B_{21} + B_{22})$
\end{itemize}
from which $C$'s block matrices can be calculated as \begin{itemize}
    \item $C_{11} = M_1 + M_4 - M_5 + M_7$,
    \item $C_{12} = M_3 + M_5$,
    \item $C_{21} = M_2 + M_4$, and
    \item $C_{22} = M_1 - M_2 + M_3 + M_6$.
\end{itemize}
This requires $7$ instead of $8$ matrix multiplications if a naive multiplication of the block matrices were used, at the cost of more additions. The same expansion can then be used on the $7$ smaller matrix multiplications.\\
This algorithm is hierarchical and contains many steps for large input matrices, but the computations themselves repeat. Also, the entire algorithm is static for a given input matrix size. Therefore, it fulfills all requirements to be implemented using CDAGs.\\
Our implementation requires $5$ different compute task types:
\begin{itemize}
    \item \texttt{Slice\{X\_SLICE, Y\_SLICE\}}: For an input matrix, this returns a slice of the matrix corresponding to \texttt{X\_SLICE} and \texttt{Y\_SLICE}. This uses valued type arguments for the slices, so no runtime overhead is created.
    \item \texttt{Add}: For two equally sized input matrices, this compute task returns their sum.
    \item \texttt{Sub}: For two equally sized input matrices, this compute task returns their difference.
    \item \texttt{MultBase}: For two equally sized input matrices, this compute task returns their multiplication, using Julia's builtin multiplication algorithm. This is used after a defined cutoff minimum size for matrices, where the Strassen algorithm is no longer reasonable.
    \item \texttt{MultStrassen}: Given the intermediate matrices $M_1 \dots M_7$, calculate and assemble $C$.
\end{itemize}
Using these, each Strassen step produces $8$ \texttt{Slice} tasks to create the four submatrices for $A$ and $B$ each, $6$ \texttt{Add} tasks, $4$ \texttt{Sub} tasks and one \texttt{StrassenMult} task. The $7$ multiplications needed are then either recursively generated as Strassen steps themselves, or as \texttt{MultBase} tasks, if the matrices are small enough.\\
Since this algorithm is only useful starting at very large matrices, and Julia already uses heavily optimized implementations for matrix multiplication by default, our implementation was not meant to be (and is not) competitive with these in terms of performance. It is, however, useful as a simple unit test and showcase of the usage of the package that does not require physics-specific knowledge.

\section{ABC-Model}\label{ch:abc_model}
\begin{figure}[t]
    \centering
    \includegraphics[width=0.8\linewidth]{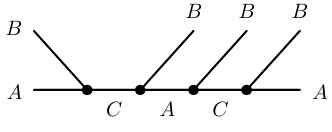}
    \caption{A tree-level Feynman diagram for the $AB \to AB^3$ scattering process in the ABC model. The inner line alternates its particle species between A-on and C-on with each B-on that binds to it.}
    \label{fig:ab_compton}
\end{figure}

The ABC-model is a QFT that is non-physical, i.e., it does not describe any real-world phenomena. It does, however, contain all required properties to be a valid QFT and can, therefore, be implemented very similarly to QED.\\
The ABC-model knows three particles, the A-on, B-on, and C-on. Similarly to QED, only one vertex exists, connecting all three particles of the model. In contrast to QED, however, none of the three particles is its own anti-particle like the photon is in QED. This means, for example, that the family of scattering processes $AB^n \to AB$ only produces valid tree-level Feynman diagrams for uneven values of $n$. As an example, Figure \ref{fig:ab_compton} shows one of the Feynman diagrams for $AB \to AB^3$.\\
Whereas the Feynman rules for QED contain mostly matrix multiplications, the computations necessary in the ABC model are only scalar multiplications. This means that while the general scaling stays the same, i.e., factorial, the matrix element computation for an ABC scattering process tends to be orders of magnitude faster than the computation of a matrix element for a QED scattering process with the same number of external particles.\\
For this reason, an implementation of this model using \texttt{ComputableDAGs.jl} is useful to compare two models with large differences in their kernel sizes but otherwise similar structure of the generated CDAGs.

\bibliographystyle{elsarticle-num}
\bibliography{bibliography.bib}

\end{document}